\documentstyle[12pt,aasms4]{article}
\begin{document}

\noindent
{\it Dissertation Summary}

\begin{center}


\title{\large \bf Stellar populations in local star-forming galaxies}

\end{center}


\author{Pablo G. P\'erez-Gonz\'alez}

\affil{Current address: University of Arizona. Steward Observatory. 933 N Cherry Av. Tucson, AZ85721, USA}

\begingroup

\parindent=1cm


\begin{center}

Electronic mail: pgperez@as.arizona.edu

Thesis work conducted at: Departamento de Astrof\'{\i}sica, Facultad de Ciencias
 F\'{\i}sicas, Universidad Complutense de Madrid (Spain)

Ph.D. Thesis directed by: J. Zamorano \& J. Gallego ;  ~Ph.D. Degree awarded: June 2003


{\it Received \underline{\hskip 5cm}}

\end{center}

\endgroup




One of the main issues in today's Astrophysics is how present day
galaxies formed and how they have evolved over time. A considerable
observational effort is being made to study galaxies from the earliest
possible times to the present. One approach to understand how
present-day galaxies came into being is to study in detail the
properties of local active star-forming galaxies, and in particular
their star-formation histories.  In this respect, it is important to
quantify the relative importance of the current episode of star
formation in comparison to the underlying older stellar
populations. 

The main goal of this thesis work \footnote{The original text is
written in Spanish and is available at
http://t-rex.fis.ucm.es/Publications/Repository/tesis\_pag\_1.ps.gz
and
http://t-rex.fis.ucm.es/Publications/Repository/tesis\_pag\_2.ps.gz}
is studying the main properties of the stellar populations embedded in
a statistically complete sample of local active star-forming galaxies:
the Universidad Complutense de Madrid (UCM) Survey of emission-line
galaxies
\footnote{http://t-rex.fis.ucm.es/UCM\_Survey/index.html}. This sample 
contains 191 local star-forming galaxies at an average redshift of
0.026. The survey was carried out using an objective-prism technique
centered at the wavelength of the H$\alpha$ nebular emission-line (a
common tracer of recent star formation).

We have compiled an extensive multi-wavelength (from the optical to
the near infrared) photometric (in narrow and broad band filters) and
spectroscopic dataset for all the galaxies in the sample. The analysis
of this dataset has been centered on the characterization of the
morphology of the objects, on the comparison of the global properties
of the sample with those of other galaxy samples at several redshifts,
and on the study of the stellar content of the UCM galaxies. In the
latter research topic, a special attention has been paid to the most
recent star-forming bursts, which are the main characteristic of our
objects. We have also developed (and tested with the dataset for the
UCM sample) a powerful stellar population synthesis tool (which is
able to obtain robust estimates of the stellar mass, a very important
parameter in the study of the evolution of galaxies), and a technique
to calculate luminosity and stellar mass functions. These techniques
may be easily used in the characterization of samples of galaxies at
any redshift.

The main results obtained in this work point out that the UCM Survey
galaxies span a broad range in properties (among others, the total
stellar mass and the star formation rate per stellar mass unit)
between those of galaxies completely dominated by current/recent star
formation (e.g., extreme dwarf HII galaxies or starburst galaxies) and
those of `normal' quiescent spirals. The UCM objects present
intermediate-late Hubble type spiral morphologies and a recent
instantaneous burst of star formation (occurred about 5 Myr ago) with
sub-solar metallicity. These bursts involved ~5\% of the total stellar
mass of the UCM galaxies. An `average' UCM galaxy has a total stellar
mass of $10^{10}$ solar masses, i.e., about a factor of 7-10 lower than
a typical local spiral galaxy. Finally, we estimate that ~$13\pm3$\% of
the total baryon mass density in the form of stars in the local
Universe is associated with star-forming galaxies.


\end{document}